\documentclass{iaus}
\usepackage[latin1]{inputenc}  
\usepackage{graphicx}

\title[Photoproduction of H$_3^+$ inside dense clouds] 
{Photoproduction of H$_3^+$ from gaseous methanol inside dense molecular
clouds}

\author[Pilling et al.]   
{S. Pilling$^1$, D. P. P. Andrade$^2$, A. C. F. Santos$^2$ \and H. M.
Boechat-Roberty$^2$}

\affiliation {$^1$LNLS, Laboratório Nacional de Luz Síncrotron, São Paulo,
Brazil. email: {\tt spilling@lnls.br} \\[\affilskip]
$^2$UFRJ, Universidade Federal do Rio de Janeiro, Rio de Janeiro, Brazil.\\
email: {\tt diana\_andrade@ufrj.br, toni@if.ufrj.br,
heloisa@ov.ufrj.br}\\[\affilskip]
}

\pubyear{2008}
\volume{IAUS251}  
 \pagerange{? -- ?}
 \date{?? and in revised form ??}
 \jname{Organic Matter in Space} \editors{A.C. Editor,
B.D. Editor \& C.E. Editor, eds.}
\begin{document}

\maketitle

\begin{abstract}
We present experimental results obtained from photoionization and
photodissociation processes of abundant interstellar methanol (CH$_3$OH) as an
alternative route for the production of H$_3^+$ in dense clouds. The
measurements were taken at the Brazilian Synchrotron Light Laboratory (LNLS)
employing soft X-ray and time-of-flight mass spectrometry. Mass spectra were
obtained using the photoelectron-photoion coincidence techniques. Absolute
averaged cross sections for the production of H$_3^+$ due to molecular
dissociation of methanol by soft X-rays (C1s edge) were determined. The
H$_3^+$'s photoproduction rate and column density were been estimated adopting
a typical soft X-ray luminosity inside dense molecular and the observed column
density of methanol. Assuming a steady state scenario, the highest column
density value for the photoproduced H$_3^+$ was about 10$^{11}$ cm$^2$, which
gives the ratio photoproduced/observed of about 0.05\%, as in the case of dense
molecular cloud AFGL 2591. Despite the small value, this represent a new and
alternative source of H$_3^+$ into dense molecular clouds and it is not been
considered as yet in interstellar chemistry models.

\end{abstract}
\firstsection 
%
\section{Introduction}

The H$_3^+$ ion plays an important role in diverse fields from chemistry to
astronomy such as, the chains of reaction that lead to the production of many
of complex molecular species observed in the interstellar medium (Herbst \&
Klemperer 1973; Dalgarno \& Black 1996; McCall \etal\ 1998 and references
therein). A detailed review about this simplest stable interstellar polyatomic
molecule could be found in Oka (2006).

In interstellar regions its main pathway formation occurs via ionization of
molecular hydrogen by the ubiquitous cosmic ray or local X-ray, followed by the
efficient ion-neutral reaction, H$_2^+ + $H$_2 \rightarrow $H$_3^+ + $H. Its
dominant destruction pathway occurs via proton-hop reaction with the abundant
interstellar carbon monoxide, H$_3^+ + $CO $\rightarrow$ HCO$^+ + $H$_2$.
However, as pointed by (Maloney, Hollenbach \& Tielens 1996 and Koyama et al.
1996) in dense clouds, mainly the ones with embedded protostars, the the soft
X-ray field may represents the dominant excitation/ionization source,
penetrating great depths into molecular clouds.

\section{Experimental methodology and results}
In an attempt to simulate the effect of stellar soft X-ray flux on gaseous
molecules inside dense clouds we have used a synchrotron radiation as a light
source. The measurements were been taken at toroidal grating monochromator
(TGM) beamline at Brazilian Synchrotron Light Laboratory (LNLS), Brazil,
employing soft X-rays photons over the C1s resonance energy range (200-310 eV).
The incoming radiation perpendicularly intersect the gas sample inside a high
vacuum chamber. Conventional time-of-flight mass spectra were obtained using
the photoelectron and photoion coincidence (PEPICO) techniques. The complete
description of the experimental setup could be found elsewhere (Boechat-Roberty
et al. 2005; Pilling et al. 2006, Pilling et al. 2007a)

Since methanol is one of the most abundant molecule in interstellar medium and
in dense molecular clouds. Therefore, even despite the reduced production of
H$_3^+$ from X-rays photodissociation process, it is reasonable to expect that
at least a fraction of the detected H$_3^+$ in molecular clouds may be produced
from this simple methyl compound molecule.

The averaged cross sections for H$_3^+$ production by soft X-rays photons at
C1s resonance, were determined taking into an account the relative intensities
of the H$_3^+$'s dissociative channels on PEPICOs spectra and the respective
simple ionization and double ionization cross section of parent molecule (see
details in Pilling et al 2007b).

The H$_3^+$'s photoproduction rate and column density were been estimated
adopting a typical soft X-ray luminosity inside dense molecular cloud (Stäuber
et al. 2005) and the observed column density of its most abundant parent ion,
methanol. The values for H$_3^+$'s photoproduction cross section due to the
dissociation of methanol by photons over the C1s edge were about 1.4 $\times$
10$^{-18}$ cm$^2$.

\section{Conclusion}

Assuming a steady state scenario and a typical X-ray luminosity of $L_x \gtrsim
10^{31}$ erg s$^{-1}$ as the case for AFGL 2591 (Stäuber et al. 2005), the
highest column density value for the photoproduced H$_3^+$ was about 10$^{11}$
cm$^2$, which gives the ratio photoproduced/observed of about 0.05\%. Despite
the small value, this represent a new and alternative source of H$_3^+$ inside
dense molecular clouds and it is not been considered as yet in interstellar
chemistry models. Better estimative for H$_3^+$ photoproduction rate depends of
more accurate soft X-ray radiation field determinations.

Moreover, the energetic ionic products released by dissociation of CH$_3$X
molecules, including the H$_3^+$ ion, become an alternative and efficient route
to complex molecular synthesis, since some ion-molecule reactions do not have
an activation barrier and are also very exothermic. We hope that these cross
section  give rise to more precise values for some molecular abundances in
interstellar clouds and even in planetary atmosphere models.




\begin{thebibliography}{}

\bibitem[Boechat-Roberty \etal\ (2005)]{Boechat2005}
{Boechat-Roberty, H.M., Pilling, S., \& Santos, A.C.F.} 2005, \textit{A\&A},
438, 915.

\bibitem[Dalgarno \& Black (1976)]{dalgarno76}
{Dalgarno, A., \& Black, J.H.} 1976, \textit{Rep. Prog. Phys.}, 39, 573.

\bibitem[Herbst \& Klemperer (1973)]{herbst73}
{Herbst, E., \& Klemperer, W.} 1973, \textit{ApJ}, 185, 505.

\bibitem[Koyama \etal\ (1996)]{Koyama96}
{Koyama, K., Hamaguchi, K., Ueno, S., Kobayashi, N., \& Feigelson E.D.} 1996,
\textit{PASJ}, 48, L87.

\bibitem[Oka \etal\ (2006)]{Oka2006}
{Oka, T.} 2006, \textit{PNAS}, 103, 12235.

\bibitem[Maloney \etal\ (1996)]{maloney96}
{Maloney, P.R., Hollenbach, D.J., \& Tielens, A.G.G.M.} 1996, \textit{ApJ},
466, 561.

\bibitem[McCall \etal\ (1998)]{mccall06}
{McCall, B.J., Geballe, T.R., Hinkle, K.H., \& Oka T.} 1998, \textit{Science},
279, 1910.

\bibitem[Stäuber \etal\ (2005)]{stauber05}
{Stäuber, P., Doty, S.D., van Dishoeck, E.F., \& Benz, A.O.} 2005,
\textit{A\&A}, 440, 949.

\bibitem[Pilling \etal\ (2006)]{Pilling2006-acetic}
{Pilling, S., Santos, A.C.F., \& Boechat-Roberty, H.M.} 2006, \textit{A\&A},
449, 1289.

\bibitem[Pilling \etal\ (2007a)]{Pilling2007a-methanol}
{Pilling, S., Neves, R., Santos, A.C.F., \& Boechat-Roberty, H.M.} 2007a,
\textit{A\&A}, 464, 393.

\bibitem[Pilling \etal\ (2007b)]{Pilling2007b-H3plus}
{Pilling, S., Andrade, D.P.P., Neves, R., Ferreira-Rodrigues, A.M., Santos,
A.C.F., \& Boechat-Roberty, H.M.} 2007b, \textit{MNRAS}, 375, 1488.

\end{thebibliography}
\end{document}